\documentclass[preprint,onecolumn,aps,pre,eqsecnum,longbibliography,showpacs,showkeys,a4paper]{revtex4-1}
\usepackage[T1]{fontenc}
\usepackage[latin9]{inputenc}
\usepackage{color}
\definecolor{note_fontcolor}{rgb}{0.80078125, 0.80078125, 0.80078125}
\usepackage[english]{babel}
\usepackage{amsmath}
\usepackage{amssymb}
\usepackage{graphicx}
\usepackage[unicode=true,
 bookmarks=false,
 breaklinks=true,pdfborder={0 0 0},backref=false,colorlinks=true]
 {hyperref}
\hypersetup{
 pdfauthor={AINS}}

\makeatletter

\newenvironment{lyxgreyedout}
  {\textcolor{note_fontcolor}\bgroup\ignorespaces}
  {\ignorespacesafterend\egroup}

\@ifundefined{textcolor}{}
{%
 \definecolor{BLACK}{gray}{0}
 \definecolor{WHITE}{gray}{1}
 \definecolor{RED}{rgb}{1,0,0}
 \definecolor{GREEN}{rgb}{0,1,0}
 \definecolor{BLUE}{rgb}{0,0,1}
 \definecolor{CYAN}{cmyk}{1,0,0,0}
 \definecolor{MAGENTA}{cmyk}{0,1,0,0}
 \definecolor{YELLOW}{cmyk}{0,0,1,0}
}
\numberwithin{equation}{section}
\numberwithin{figure}{section}
\numberwithin{table}{section}

\usepackage{graphicx}

\DeclareGraphicsExtensions{.png,.pdf,.eps}
\graphicspath{{pictures/}}

\@ifundefined{showcaptionsetup}{}{%
 \PassOptionsToPackage{caption=false}{subfig}}
\usepackage{subfig}
\makeatother

\begin{document}

\preprint{}

\title{Extreme beam attenuation in double-slit experiments: Quantum and
subquantum scenarios\vspace{1cm}
}

\author{Gerhard \surname{Grössing}\textsuperscript{1}}

\author{Siegfried \surname{Fussy}\textsuperscript{1}}

\author{Johannes \surname{Mesa Pascasio}\textsuperscript{1,2}}

\author{Herbert \surname{Schwabl}\textsuperscript{1}}

\email[Corresponding author: ]{ains@chello.at}

\selectlanguage{english}%

\affiliation{\textsuperscript{1}Austrian Institute for Nonlinear Studies, Akademiehof,
Friedrichstr.~10, 1010 Vienna, Austria}

\affiliation{\textsuperscript{2}Atominstitut, TU Wien, Operng.~9, 1040 Vienna,
Austria\vspace{2cm}
}

\date{\today}
\begin{abstract}
Combining high and low probability densities in\emph{ intensity hybrids},
we study some of their properties in double-slit setups. In particular,
we connect to earlier results on beam attenuation techniques in neutron
interferometry and study the effects of very small transmission factors,
or very low counting rates, respectively, at one of the two slits.
We use a ``superclassical'' modeling procedure which we have previously
shown to produce predictions identical with those of standard quantum
theory. Although in accordance with the latter, we show that there
are previously unexpected new effects in intensity hybrids for transmission
factors below $a\lesssim10^{-4}$, which can eventually be observed
with the aid of weak measurement techniques. We denote these as \textsl{quantum
sweeper} effects, which are characterized by the bunching together
of low counting rate particles within very narrow spatial domains.
We give an explanation of this phenomenology by the circumstance that
in reaching down to ever weaker channel intensities, the nonlinear
nature of the probability density currents becomes ever more important,
a fact which is generally not considered -- although implicitly present
-- in standard quantum mechanics.%
\begin{lyxgreyedout}
\noindent \global\long\def\VEC#1{\mathbf{#1}}
\global\long\def\d{\,\mathrm{d}}
\global\long\def\e{{\rm e}}
\global\long\def\meant#1{\left<#1\right>}
\global\long\def\meanx#1{\overline{#1}}
\global\long\def\mpbracket{\ensuremath{\genfrac{}{}{0pt}{1}{-}{\scriptstyle (\kern-1pt +\kern-1pt )}}}
\global\long\def\pmbracket{\ensuremath{\genfrac{}{}{0pt}{1}{+}{\scriptstyle (\kern-1pt -\kern-1pt )}}}
\global\long\def\p{\partial}
\end{lyxgreyedout}

\end{abstract}


\maketitle

\section{Introduction\label{sec:Introduction}}

Continuing the search for new, and perhaps surprising, features of
quantum systems, one option is to steadily decrease the intensity
of a channel in one spatially constrained area, as compared to a reference
intensity in another, equally constrained area. For example, one can
employ the usual double-slit experiments and modify one of the two
slits' channels such that the corresponding outgoing probability density
is very low compared to that of the other slit. We call a combination
of such distributions of high and low probability densities, or intensities,
respectively,\emph{ intensity hybrids}.

Since the 1980ies, one possibility to experimentally establish and
probe such hybrids has been through the introduction of beam attenuation
techniques, as demonstrated in the well-known papers by Rauch's group
in neutron interferometry~\citep{Rauch.1984static,Rauch.1990low-contrast}.
In the present paper, we re-visit these experiments and results from
a new perspective, and we also discuss new, previously unexpected
effects. For, our group has in recent years introduced a ``superclassical''
approach to describe and explain quantum behavior as an emergent phenomenon
in between classical boundary conditions on the one hand, and an assumed
classical subquantum domain at vastly smaller spatial scales on the
other~\citep{Groessing.2008vacuum,Groessing.2009origin,Groessing.2010emergence,Groessing.2011dice,Groessing.2011explan,Groessing.2012doubleslit,Groessing.2013dice,Fussy.2014multislit,Groessing.2014relational}.
Here we are going to apply our approach to the above-mentioned intensity
hybrids. Our main result is that in employing ever weaker channel
intensities, nonlinear effects become ever more important, which are
generally not considered -- although implicitly present -- in ordinary
quantum mechanics, but which are a crucial characteristic of subquantum
models as the one developed by our group. Experimental tests are feasible
with the aid of weak measurement techniques.

\section{Deterministic and stochastic beam attenuation in the double slit
and their superclassical modeling\label{sec:BeamAttenuation}}

\subsection*{2.1 Beam attenuation in neutron interferometry\label{sub:2.1 neutron interferometry}}

Deterministic and stochastic beam attenuation have been studied extensively
in neutron interferometry, beginning with the work by Rauch and Summhammer
in 1984~\citep{Rauch.1984static}. More recently, an interesting
model of these results has been presented by De~Raedt \emph{et al}.~\citep{DeRaedt.2012discrete-event}
with the aid of event-by-event simulations, thus confirming the possibility
to describe the known results even without the use of quantum mechanics.
Our approach, in contrast, though also not relying on the quantum
mechanical formalism, nevertheless is an attempt at a ``deeper level''
modeling from which the quantum mechanical results are expected to
emerge. In other words, as our model is intended ultimately to go
beyond standard quantum mechanics, but also to provide the quantum
results as a limiting case, we shall first use the physics of beam
attenuation as a means to verify the agreement with the quantum mechanical
predictions. In a second step, then, we shall exploit the ``extremes''
of the quantum as well as of our superclassical descriptions, respectively,
i.e.\ consider parameter values that cover a vast range of orders
of magnitude so as to study extreme examples of intensity hybrids
as introduced in the first Section. We shall then find results that
look rather surprising from the viewpoint of standard quantum mechanics,
but are fully understandable with our model.

In~\citep{Rauch.1984static,Rauch.1990low-contrast}, a beam chopper
was used as a deterministic absorber in one arm of a two-armed interferometer,
whereas for stochastic absorption semitransparent foils of various
materials were used. Despite the net effect of the same percentage
of neutrons being attenuated, the quantum mechanical formalism predicts
the following different behaviors for the two cases. Introducing the
\emph{transmission factor} $a$ as the beam's transmission probability,
in the case of a (deterministic) chopper wheel it is given by the
temporal open-to-closed ratio, $a=\frac{t_{open}}{t_{open}+t_{closed}}$
, whereas for a (stochastic) semitransparent material defined by its
absorption cross section, it is simply the relation of the intensity
$I$ with absorption compared to the intensity $I_{0}$ without, i.e.\ $a=I/I_{0}$.
Thus the beam modulation behind the interferometer is obtained in
the following two forms. For the deterministic chopper system the
intensity is, with $\varphi$ denoting the phase difference, given
by
\begin{equation}
I\propto\left(1-a\right)\left|\varPsi_{1}\right|^{2}+a\left|\varPsi_{1}+\varPsi_{2}\right|^{2}\propto1+a+2a\cos\varphi,\label{eq:sw.1}
\end{equation}
whereas for stochastic beam attenuation with the semitransparent material
it is
\begin{equation}
I\propto\left|\varPsi_{1}+\varPsi_{2}\right|^{2}\propto1+a+2\sqrt{a}\cos\varphi.\label{eq:sw.2}
\end{equation}
In other words, although the same number of neutrons is observed in
both cases, in the first one the contrast of the interference pattern
is proportional to $a$, whereas in the second case it is proportional
to $\sqrt{a}$.

In our accounting for the just described attenuation effects, we choose
the usual double slit scenario, primarily because this will be very
useful later on when discussing more extreme intensity hybrids. However,
before describing the results a brief review is in order to indicate
the essentials of our model which are employed in our explanation
of the beam attenuation phenomena.

\subsection*{2.2 Short review of the superclassical approach to quantum mechanics}

Throughout the last years we have developed an approach to quantum
mechanics within the scope of theories on ``Emergent Quantum Mechanics''.
(For the proceedings of the first two international conferences devoted
to this subject, see~\citep{Groessing.2012emerqum11-book,Groessing.2014emqm13-book}.)
Essentially, we consider the quantum as a complex dynamical phenomenon,
rather than as representing some ultimate-level phenomenon in terms
of, e.g., pure formalism, wave mechanics, or strictly particle physics
only. Our assumption is that a particle of energy $E=\hbar\omega$
is actually an oscillator of angular frequency $\omega$ phase-locked
with the zero-point oscillations of the surrounding environment, the
latter of which containing both regular undulatory and fluctuating
components and being constrained by the boundary conditions of the
experimental setup via the emergence of standing waves. In other words,
the particle in this approach is an off-equilibrium steady-state maintained
through the ``sub-quantum arrow of time'', in T. Nieuwenhuizen's~\citep{Nieuwenhuizen.2014subquantum}
terminology, i.e.\ by the throughput of zero-point energy from its
vacuum surroundings. This is in close analogy to the bouncing/walking
droplets in the experiments of Couder and Fort's group~\citep{Fort.2010path-memory,Couder.2006single-particle,Couder.2012probabilities},
which in many respects can serve as a classical prototype guiding
our intuition. However, we denote our whole ansatz as ``superclassical''~\citep{Groessing.2014emqm13-book},
because it connects the classical physics at vastly different scales,
i.e.\ the ordinary classical one and an assumed subquantum one, with
``new'' effects emergent on intermediate scales, which we have come
to know and describe as quantum ones.

In fact, throughout the years we have succeeded in reproducing a number
of quantum mechanical results with our superclassical model, i.e.\ without
any use of the quantum mechanical formalism, like states, wave functions,
\textit{et cetera}. Note, moreover, that a Gaussian emerging from,
say, a single slit with rounded edges (so as to avoid diffraction
effects) is in our model the result of statistically collecting the
effects of the aleatory bouncing of our particle oscillator, and thus
not an ontic entity \emph{per se}. Rather, the Gaussian stands for
the statistical mean of the ``excitation'' (or ``heating up'')
of the medium within the confines of the slit, and later, as the bouncer/walker
progresses, further away from it. We have described this in terms
of a thermal environment that represents stored kinetic energy in
the vacuum and that is responsible for where the particle is being
guided to. For example, consider particle propagation coming out from
a Gaussian slit. For a particle exactly at the center of the Gaussian,
the diffusive momentum contributions from the heated up environment
will on average cancel each other for symmetry reasons. However, the
further off the particle is from that center, the stronger the symmetry
will be broken, thus resulting in a position-dependent net acceleration
or deceleration, respectively -- in effect, resulting in the decay
of the wave packet. (For a detailed analysis, see~\citep{Groessing.2010emergence}.)
In other words, due to wave-like diffusive propagations originating
from the particle's bounces to stir up the medium of the vacuum, particle
paths can be influenced by the agitations of the vacuum even in places
where no other particle is around. This will be important for the
discussion of intensity hybrids later on.

The essential features of our model can be summarized as follows.
As already mentioned, we have shown that the spreading of a wave packet
can be exactly described by combining the forward (convective) with
the orthogonal diffusive velocity fields. The latter fulfill the condition
of being unbiased w.r.t.\ the convective velocities, i.e.~the orthogonality
relation for the \textit{averaged} velocities derived in~\citep{Groessing.2010emergence}
is $\VEC{\overline{vu}}=0$, since any fluctuations $\VEC u=\delta\left(\nabla S/m\right)$
are shifts along the surfaces of action $\mathit{S=\mathrm{\mathrm{const}}.}$
Moreover, the fluctuations can be directed towards the left or towards
the right from the mean (i.e.\ from the Ehrenfest trajectory), which
leads us to introduce the notations $\mathbf{u}_{iL}$ and $\mathbf{u}{}_{iR}$,
respectively.

Reducing the general case discussed in \citep{Fussy.2014multislit}
to the double-slit case, we note for the first and second channels
the emergent velocity vectors $\VEC v_{\mathrm{1(2)}},\VEC u_{\mathrm{1(2)R}},$
and $\VEC u_{\mathrm{1(2)L}}$ with associated amplitudes $R_{1(2)}$,
respectively. In order to completely accommodate the totality of the
system of currents present, we obtain a local wave intensity for any
velocity component, e.g.\ for $\VEC v_{\mathrm{1}}$, by the pairwise
projection on the unit vector $\VEC{\hat{v}}_{1}$ weighted by $R_{1}$
of the totality of all amplitude weighted unit velocity vectors being
operative at $\mathrm{(}\VEC x,t)$:
\begin{equation}
P(\VEC v_{\mathrm{1}})=R_{1}\VEC{\hat{v}}_{1}\cdot(\VEC{\hat{v}}_{1}R_{1}+\VEC{\hat{u}_{\mathrm{1R}}\mathrm{\mathit{R_{\mathrm{1}}}}}+\VEC{\hat{u}_{\mathrm{1L}}\mathrm{\mathit{R_{\mathrm{1}}}}}+\VEC{\hat{v}}_{2}R_{2}+\VEC{\hat{u}_{\mathrm{2R}}\mathit{R_{\mathrm{2}}}}+\VEC{\hat{u}}_{\mathrm{2L}}R_{2}).\label{eq:sw.3}
\end{equation}
Due to the mentioned orthogonality between $\VEC v_{i}$ and $\mathbf{u}_{iL(R)}$,
i.e.\ with $\VEC{\hat{v}}_{1}\cdot\VEC{\hat{v}}_{2}=:\cos\varphi$,
but $\VEC{\hat{v}}_{i}\cdot\VEC{\hat{u}}_{iL(R)}=0$ \textit{et cetera},
and with the trivial relation $\VEC{\hat{u}}_{i\mathrm{R}}R_{i}+\VEC{\hat{u}}_{i\mathrm{L}}R_{i}=0$,
$i=1,2$, we finally obtain
\begin{align}
P(\VEC v_{\mathrm{1}}) & =R_{1}^{2}+R_{1}R_{2}\cos\varphi\label{eq:sw.4}\\
P(\VEC u_{\mathrm{1R}}) & =-P(\VEC u_{\mathrm{1L}})=\sin\varphi\\
P(\VEC v_{\mathrm{2}}) & =R_{2}^{2}+R_{1}R_{2}\cos\varphi\\
P(\VEC u_{\mathrm{2R}}) & =-P(\VEC u_{\mathrm{2L}})=-\sin\varphi
\end{align}
leading to the probability densities $P_{i}$ for each channel, $i=1$
or $2$, 
\begin{align}
P_{1} & =R_{1}^{2}+R_{1}R_{2}\cos\varphi\label{eq:sw.8}\\
P_{2} & =R_{2}^{2}+R_{1}R_{2}\cos\varphi.\label{eq:sw.9}
\end{align}

The total probability density $P_{\mathrm{tot}}$ for the double-slit
case is then simply given by $P_{\mathrm{tot}}=P_{1}+P_{2}.$ The
local current attributed to each velocity component is defined as
the corresponding ``local'' intensity-weighted velocity, e.g.\ for
$\VEC v_{1}$ it is given as $\VEC J\mathrm{(}\VEC v_{\mathrm{1}})=\VEC v_{\mathrm{1}}P(\VEC v_{\mathrm{1}})=\VEC v_{\mathrm{1}}\left(R_{1}^{2}+R_{1}R_{2}\cos\varphi\right)$.
The local intensity of a partial current is dependent on all other
currents, and the total current itself is composed of all partial
components, thus constituting a representation of what we call \emph{relational
causality}. The total current consequently reads as $\VEC J_{\mathrm{tot}}=\VEC v_{\mathrm{1}}P(\VEC v_{\mathrm{1}})+\VEC u_{1R}P\mathrm{(}\VEC u_{1\mathrm{R}}\mathrm{)}+\VEC u_{1\mathrm{L}}P\mathrm{(}\VEC u_{\mathrm{1\mathrm{L}}}\mathrm{)}+\VEC v_{\mathrm{2}}P(\VEC v_{\mathrm{2}})+\VEC u_{2R}P\mathrm{(}\VEC u_{2\mathrm{R}}\mathrm{)}+\VEC u_{\mathrm{2L}}P\mathrm{(}\VEC u_{2\mathrm{L}}\mathrm{)}$,
which, by identifying the resulting diffusive velocities $\VEC u_{i\mathrm{\mathrm{R}}}-\VEC u_{i\mathrm{L}}$
with the effective diffusive velocities $\VEC u_{i}$ for each channel,
finally leads to
\begin{equation}
\VEC J_{\mathrm{tot}}=R_{1}^{2}\VEC v_{\mathrm{1}}+R_{2}^{2}\VEC v_{\mathrm{2}}+R_{1}R_{2}\left(\VEC v_{\mathrm{1}}+\VEC v_{2}\right)\cos\varphi+R_{1}R_{2}\left(\VEC u_{1}-\VEC u_{2}\right)\sin\varphi.\label{eq:sw.10}
\end{equation}
The trajectories or streamlines, respectively, are given by
\begin{equation}
\VEC{\dot{x}}=\VEC v_{\mathrm{tot}}=\frac{\VEC J_{\mathrm{tot}}}{P_{\mathrm{tot}}}\thinspace.\label{eq:sw.11}
\end{equation}
As first shown in~\citep{Groessing.2012doubleslit}, by re-inserting
the expressions for convective and diffusive velocities, respectively,
i.e.\ 
\begin{equation}
\VEC v_{i}=\frac{\nabla S_{i}}{m},\quad\textrm{ and }\quad\VEC u_{i}=-\frac{\hbar}{m}\frac{\nabla R_{i}}{R_{i}}\,,\label{eq:sw.12}
\end{equation}
one immediately identifies Eq.~(\ref{eq:sw.11}) with the Bohmian
guidance equation and Eq.~(\ref{eq:sw.10}) with the quantum mechanical
pendant for the probability density current~\citep{Sanz.2008trajectory}.
As we have shown also the latter identity, we are assured that our
results are the same as those of standard quantum mechanics -- provided,
of course, that generally (i.e.\ in the standard quantum as well
as in our ansatz) the idealization of using Gaussians or similar regular
distribution functions is applicable for the high degrees of attenuation
studied here. However, as in our model we can also make use of the
velocity field to plot the averaged particle trajectories, we can
in principle provide a more detailed picture, in similar ways to the
Bohmian one, but still not relying on any quantum mechanical tool
like a wave function, for example.

\subsection*{2.3 Application to deterministic and stochastic beam attenuation
experiments}

Let us now display some typical results from our superclassical approach
to beam attenuations. We can classically simulate the propagation
of a Gaussian whose variance increases due to the diffusion process
we implement~\citep{Groessing.2010emergence}. The results are in
perfect agreement with the quantum mechanical ones. This can be seen
already from the resulting formulas for probability density distributions,
which, despite a different approach, lead to the same final forms.

\begin{figure}[!tbh]
\subfloat[$a=0.25$.\label{fig:sw.1a}]{\centering{}\includegraphics[bb=0bp 0bp 240bp 198bp]{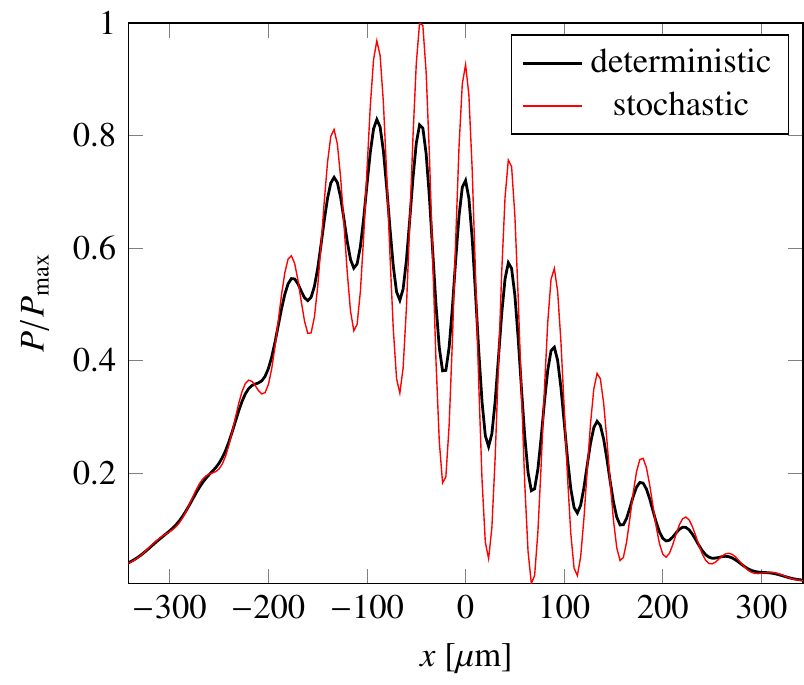}}\subfloat[$a=0.025$. \label{fig:sw.1b}]{\centering{}\includegraphics[bb=0bp 0bp 240bp 198bp]{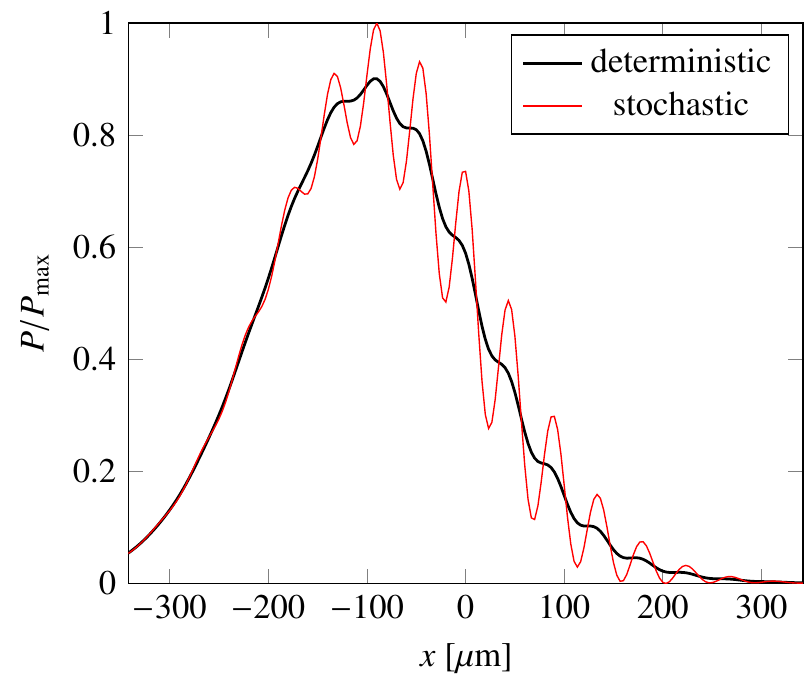}}

\subfloat[$a=0.0025$. \label{fig:sw.1c}]{\centering{}\includegraphics[bb=0bp 0bp 240bp 198bp]{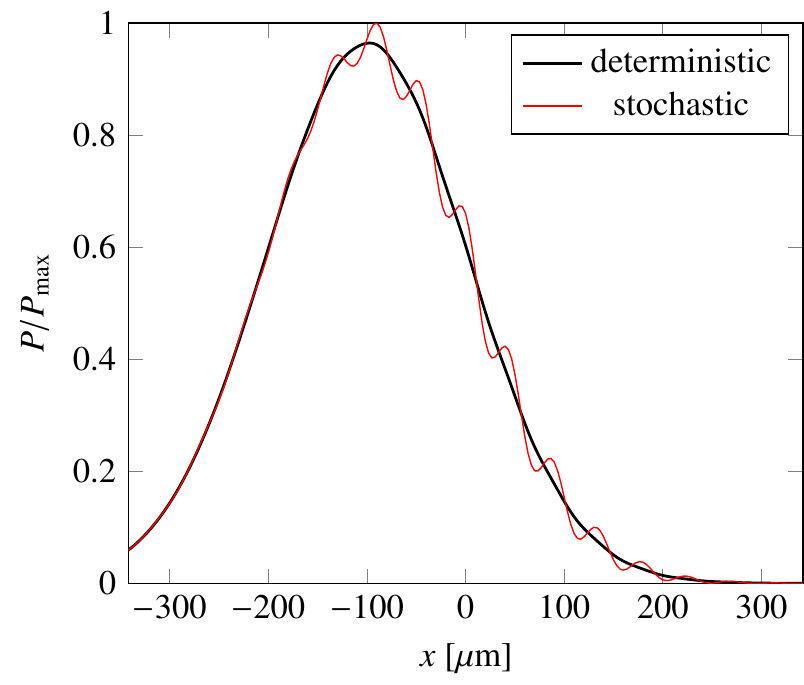}}

\protect\caption{Superclassical simulation of probability density distributions with
beam attenuation $a$ at slit~2.\label{fig:sw.1}}
\end{figure}

To begin, consider deterministic attenuation first. This case is also
very straightforward to understand in our approach, as it just amounts
to the addition of two context-dependent probability distributions,
the first context being a single-slit experiment, and the second context
a double-slit one. For the calculation, we just have to use the fact
that in our model the intensity according to Eqs.~(\ref{eq:sw.8})
and~(\ref{eq:sw.9}) is given by the projection rule $P_{i}=R_{i}^{2}+R_{i}R_{j}\cos\varphi$
for the real-valued wave amplitudes $R_{i}$ per slit~$i$  with
$j$ denoting the second slit. Keeping in mind that in the one-slit
case as temporarily realized in the deterministic beam chopper experiment,
the density of the single open slit is simply given as $P_{1}^{'}=R_{1}^{2}$~\citep{Fussy.2014multislit},
we have in complete agreement with~(\ref{eq:sw.1}) that
\begin{equation}
I\propto\left(1-a\right)P_{1}^{'}+a\left(P_{1}+P_{2}\right)\propto1+a+2a\cos\varphi.
\end{equation}
On the other hand, for stochastic attenuation we obtain with $a=I/I_{0}$,
and thus with the amplitude of the attenuated slit (e.g.\ number
$2$) becoming $R_{2}\rightarrow\sqrt{a}R_{2}$ , that
\begin{equation}
I\propto\left(P_{1}+P_{2}\right)\propto1+a+2\sqrt{a}\cos\varphi,
\end{equation}
which is in complete agreement with~(\ref{eq:sw.2}). In Figs.~\ref{fig:sw.1a}--\ref{fig:sw.1c}
we show the results of our computer simulations following~\citep{Groessing.2010emergence}
for the probability density distributions of a neutron beam for three
different values of the beam transmission factor $a$. The typical
wavelength used is $\lambda=1.8\thinspace\mathrm{nm}$ (c.f.~\citep{Rauch.2000neutron}).
The Gaussian slits each are $22\thinspace\mu\mathrm{m}$ wide, with
their centers being $200\thinspace\mu\mbox{m}$ apart, and the intensity
distributions are recorded on a screen located in the forward direction
at a distance of $5\thinspace\mbox{m}$ from the double slit. Corresponding
to the different behaviors of the contrast in deterministic and stochastic
attenuation, respectively, one can see the different contributions
to the overall probability density distribution, with the differences
becoming smaller and smaller with ever decreasing transmission factor
$a$. For consistency, we have also checked and confirmed that the
total areas below the two curves are identical, as they must be in
order to represent the same overall throughput of the number of neutrons.

In interpreting their results of the beam attenuation experiments
with neutrons, the authors accordingly found evidence in support of
the complementarity principle. That is, the more pronounced the visibility
of the interference fringes, the less which-path knowledge one can
have of the particle propagation, and \emph{vice versa}: the higher
the probability is for a particle to take a path through one certain
slit, the less visible the interference pattern becomes. This was
in fact confirmed in the above-mentioned neutron interferometry experiments,
albeit to a lesser degree for very low counting rates. In particular,
the authors of~\citep{Rauch.1984static,Rauch.1990low-contrast} often
use expressions such as the ``particle-like'' or the ``wave-like''
nature of the quantum system studied, depending on whether which-path
information or interference effects are dominant, respectively. While
this is all correct as far as the mentioned papers are concerned,
an extrapolation of the use of ``particle-like'' or ``wave-like''
attributed to more extreme intensity hybrids is not guaranteed. In
fact, we shall show in the following Section a particular effect which
undermines said dichotomy of ``particle-like'' and ``wave-like''
features, thereby calling for an improved, more general analysis of
possible relationships between particle and wave features.

\section{Phenomenology of the quantum sweeper for coherent and incoherent
beams\label{sec:quantum sweeper}}

We assume a coherent beam in a double-slit experiment, with the intensity
distribution being recorded on a screen, and we are going to discuss
a particular effect of the stochastic attenuation of one of the two
emerging Gaussians at very small transmission factors. With the appropriate
filtering of the particles going through one of the two slits, the
recorded probability density on a screen in the surroundings of the
experiment will appear differently than what one would normally expect.
That is, even if one had a low beam intensity coming from one slit,
one would expect the following scenario according to the usual quantum
mechanical heuristics: The interference pattern would more and more
become asymmetric in the sense that the contributions from the fully
open slit would become dominant until such a low counting rate from
the attenuated slit is arrived at that essentially one would have
a one-slit distribution of recorded particles on the screen. This
tendency is at least clearly visible in Figs.~\ref{fig:sw.1a}--\ref{fig:sw.1c},
and one would only expect for ever smaller values of $a$ that the
oscillatory behavior of the stochastic case would more and more disappear
to finally merge with the smoothed-out pattern of an essentially one-slit
distribution pattern.

\begin{figure*}[!tbh]
\begin{centering}
\subfloat[$a=10^{-1}$\label{fig:sw.2a}]{\begin{centering}
\includegraphics[width=0.39\textwidth]{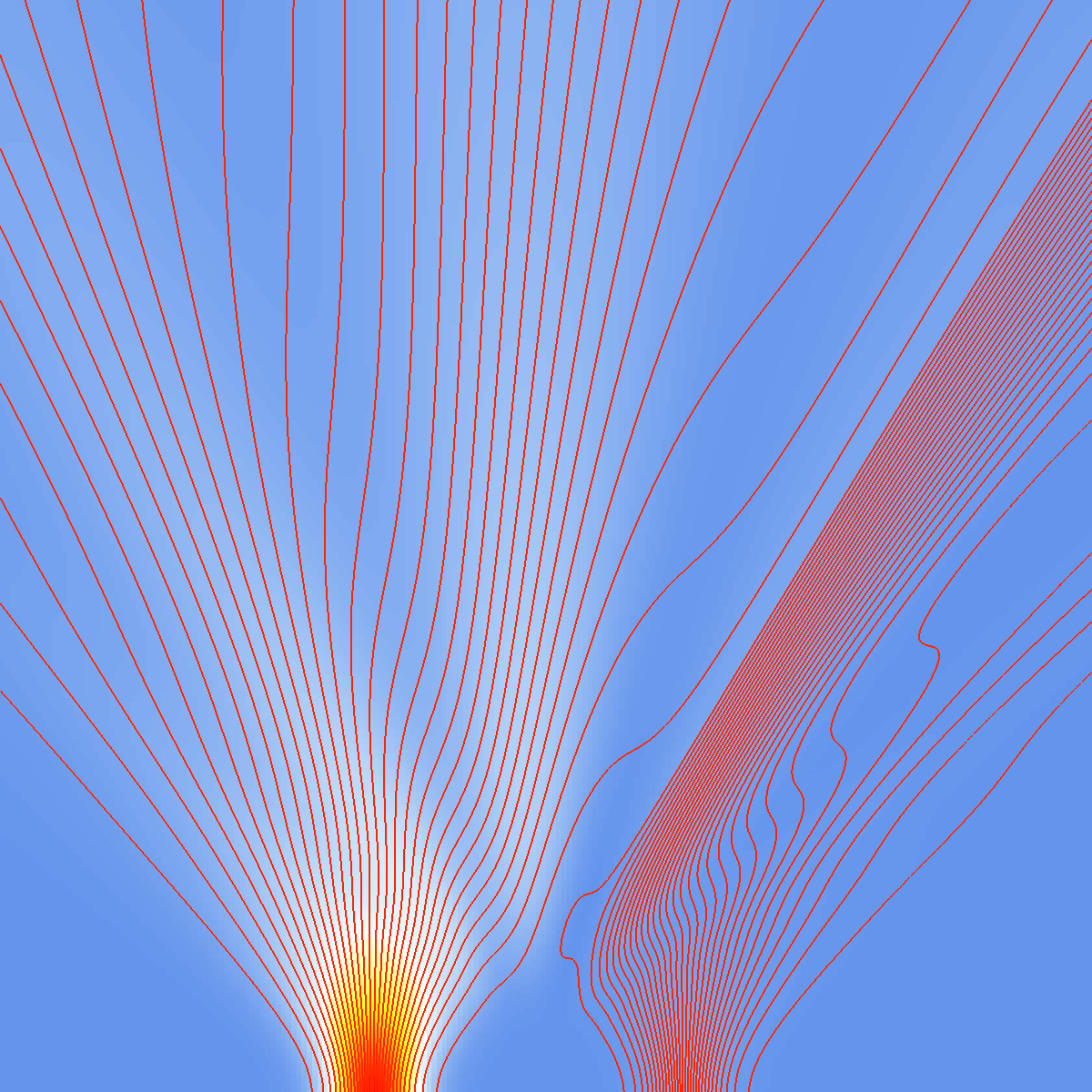} 
\par\end{centering}

}\subfloat[$a=10^{-2}$\label{fig:sw.2b}]{\begin{centering}
\includegraphics[width=0.39\textwidth]{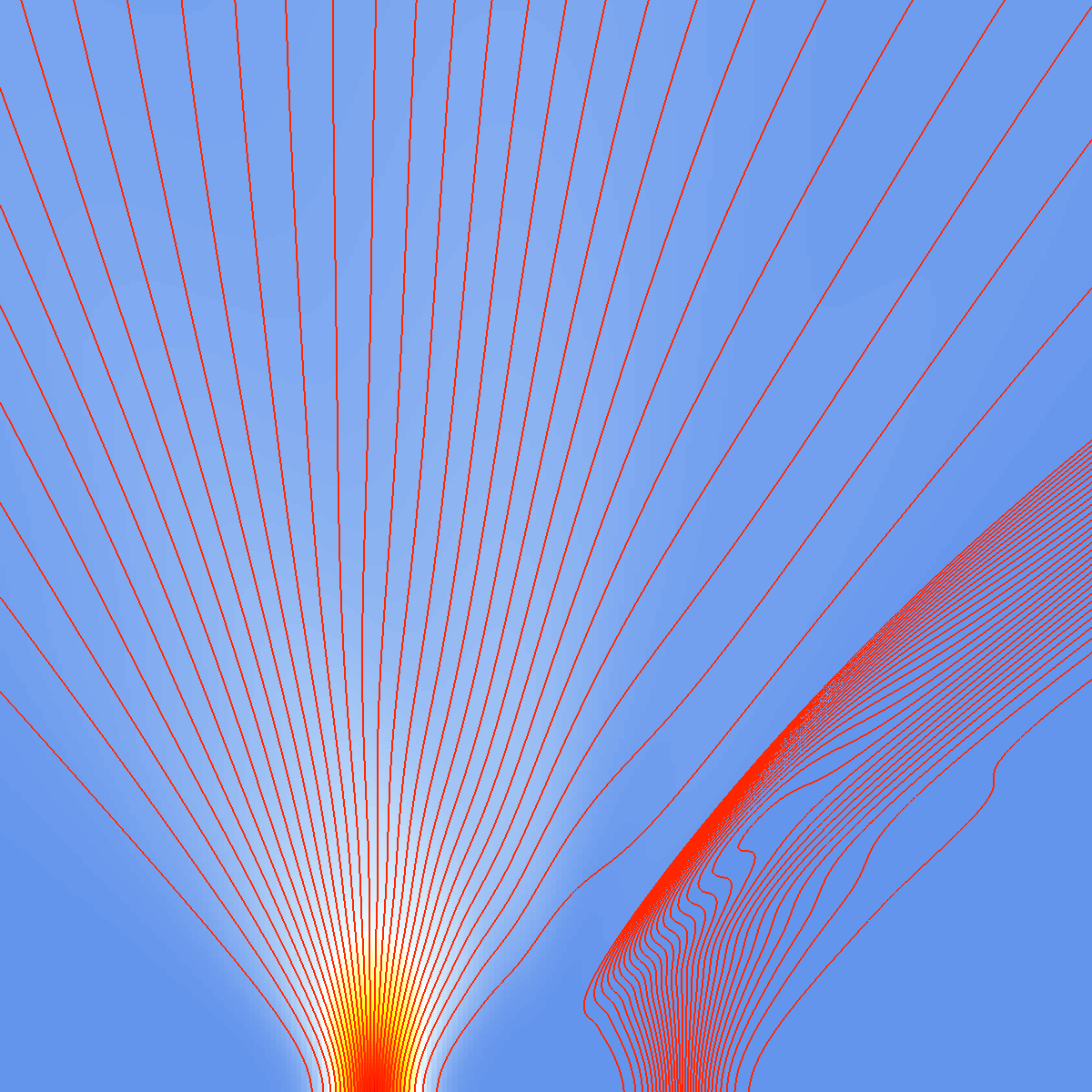}
\par\end{centering}

}
\par\end{centering}

\begin{centering}
\subfloat[$a=10^{-4}$\label{fig:sw.2b-1}]{\begin{centering}
\includegraphics[width=0.39\textwidth]{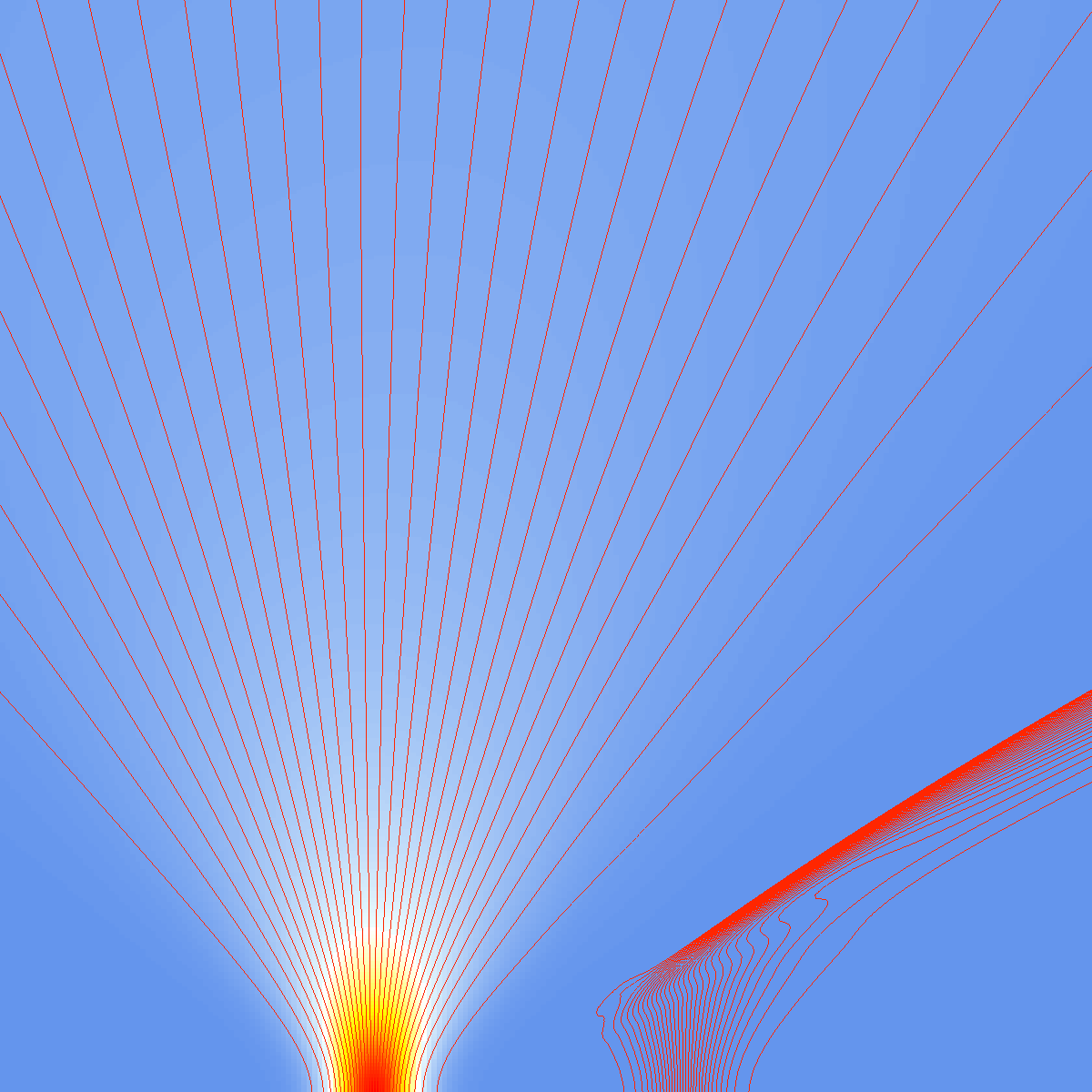}
\par\end{centering}

}\subfloat[$a=10^{-10}$\label{fig:sw.2c}]{\begin{centering}
\includegraphics[width=0.39\textwidth]{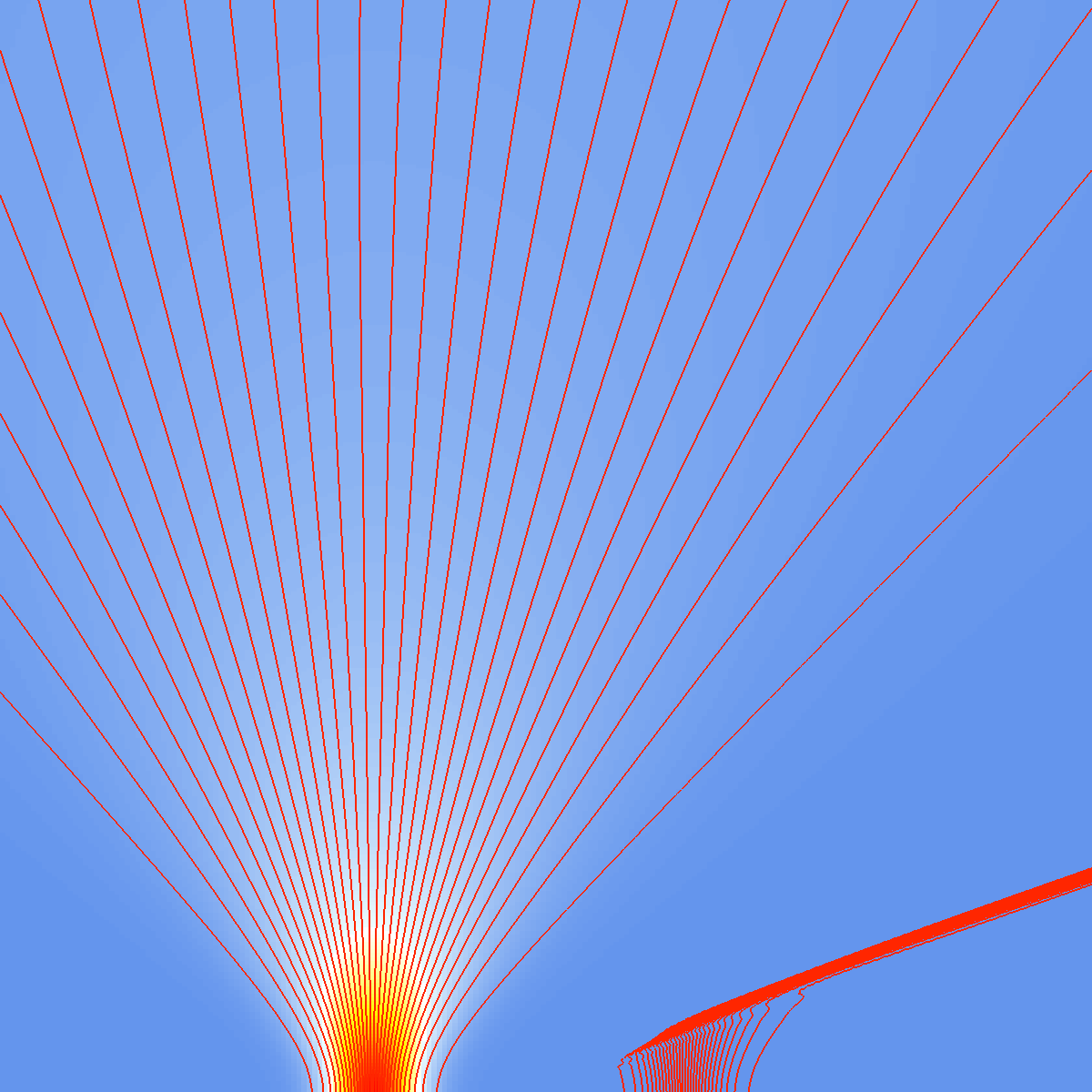}
\par\end{centering}

}
\par\end{centering}

\protect\caption{The ``quantum sweeper'' effect for different transmission factors
$a$: With ever lower values of $a$, one can see a steadily growing
tendency for the low counting rate particles of the attenuated beam
to become swept aside. In our model, this phenomenology is explained
by processes of diffusion, due to the presence of accumulated vacuum
heat (i.e.\ kinetic energy) mainly in the ``strong'' beam. The
sweeper effect is thus the result of the vacuum heat sweeping aside
the very low intensity beam, with a ``no crossing'' line defined
by the balancing out of the osmotic momenta coming from the two beams,
respectively. Throughout this paper, to demonstrate the effect more
clearly, we use the same number of trajectories for each slit.\label{fig:sw.2}}
\end{figure*}

Interestingly, this is not exactly what one obtains at least for very
low values of $a$ when going through the calculations and/or computer
simulations with our superclassical bouncer model. The latter consists,
among other features, in an explicit form of the velocity field emerging
from the double slit, as well as of the probability density current
associated with it. 

Fig.~\ref{fig:sw.2} shows the ``quantum sweeper'' effect: a series
of probability density distributions plus averaged trajectories for
the case that the intensity in slit~2 is gradually diminished. We
use the same model as in~\citep{Groessing.2012doubleslit}, or \citep{Holland.1993},
respectively: particles (represented by plane waves in the forward
$y$-direction) from a coherent source passing through ``soft-edged''
slits in a barrier (located along the $x$-axis) and recorded at a
screen in the forward direction, i.e.\ parallel to the barrier. This
situation is described by two Gaussians representing the totality
of the effectively ``heated-up'' path excitation field, one for
slit~1 and one for slit~2, whose centers have the same distances
from the plane spanned by the source and the center of the barrier
along the $y$-axis, respectively. Now, with ever lower values of
the transmission factor $a$ during beam attenuation, one can see
a steadily growing tendency for the low counting rate particles of
the attenuated beam to become swept aside. In our model, this is straightforward
to understand, because we have the analytical tools to differentiate
between the forward propagations $\VEC v_{i}$ and the diffusive influences
of velocities $\VEC u_{i}$, as distinguishable contributions from
the different slits~$i$. Thus, it is processes of diffusion which
are seen in operation here, due to the presence of accumulated heat
(i.e.\ kinetic energy), primarily in the ``strong'' beam, as discussed
in the previous Section. So, in effect, we understand Fig.~\ref{fig:sw.2}
as the result of the vacuum heat sweeping aside the very low intensity
beam, with a ``no crossing'' line defined by the balancing out of
the diffusive momenta, $m\left(\VEC u_{1}+\VEC u_{2}\right)=0.$ 

\begin{figure*}[!tbh]
\begin{centering}
\includegraphics[width=1\textwidth]{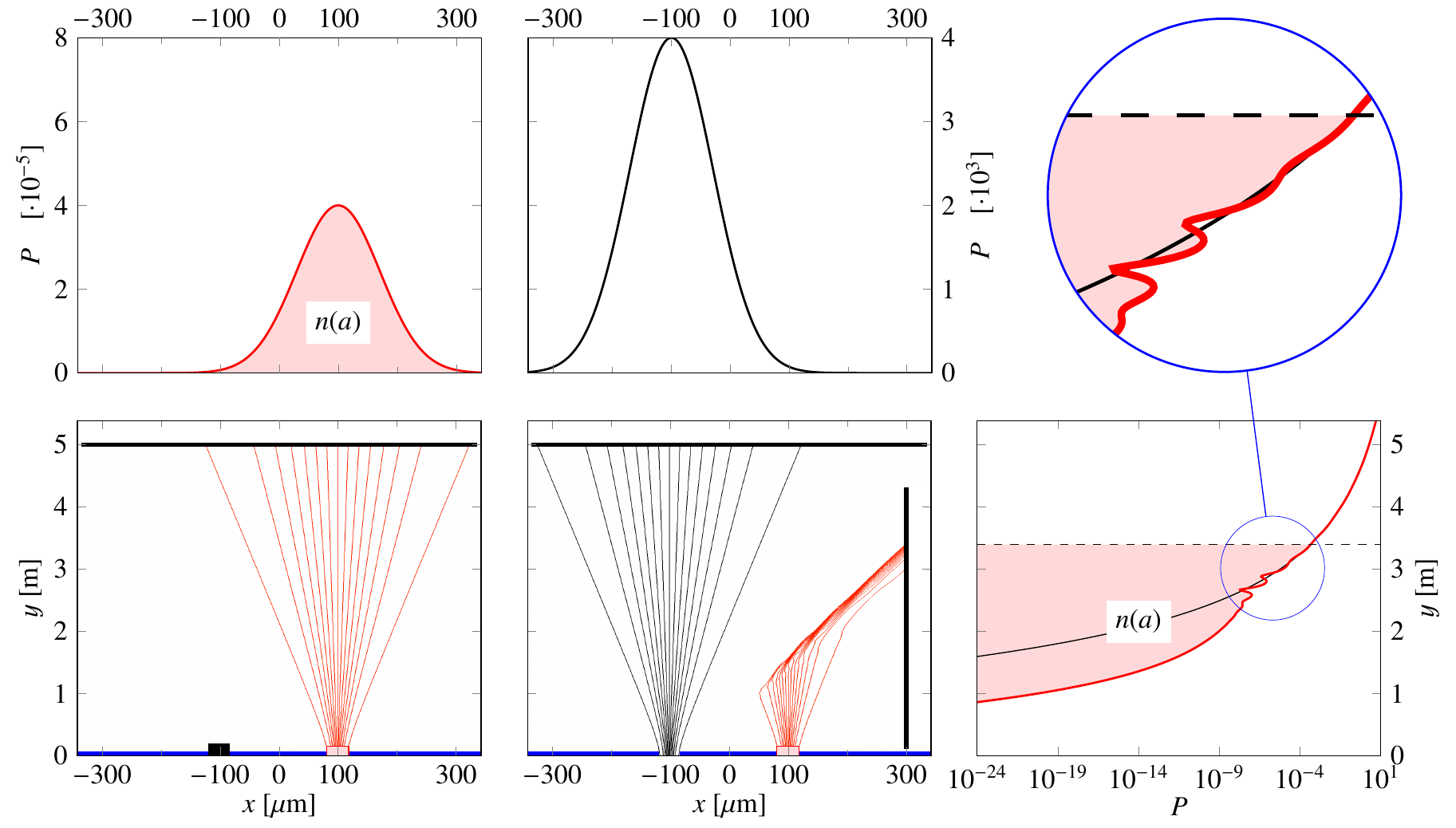}
\par\end{centering}

\begin{centering}
\hfill{}\subfloat[\label{fig:sw.4a}]{}\hfill{}\hfill{}\subfloat[\label{fig:sw.4b}]{}\hfill{}\hfill{}\subfloat[\label{fig:sw.4c}]{}\hfill{}
\par\end{centering}

\protect\caption{Registration of particles during extreme beam attenuation, $a=10^{-8}$.
(a) If slit~1 is closed, a small number $n\left(a\right)$ of particles
coming from slit~2 is registered on the forward screen. (b) If, then,
slit~1 is fully opened (i.e.\ with $a=1$), one registers a much
higher number of particles for slit~1, but apparently none for slit~2.
Instead, $n\left(a\right)$ particles are then registered on the sideways
screen parallel to the $y$-axis. (c) The probability density distribution
for the latter exhibits marked signs of interference effects due to
the compressed wave superpositions within the bunching area caused
by the sweeper effect.\label{fig:sw.4}}
\end{figure*}

Importantly, for certain slit configurations and sizes of the transmission
factor, the sweeper effect leads to a bunching of trajectories which
may become deflected into a direction almost orthogonal to the original
forward direction. In other words, one would need much wider screens
in the forward direction to register them, albeit then weakened due
to a long traveling distance. On the other hand, if one installed
a screen orthogonal to the ``forward screen'', i.e.\ one that is
parallel to the original forward motion (and thus to the $y$-axis),
one could significantly improve the contrast and thus register the
effect more clearly (see also Fig.~\ref{fig:sw.4} further below).
Further, we note that changing the distance between the two slits
does not alter the effect, but demonstrates the bunching of the low
counting rate arrivals in essentially the same narrow spatial area
even more drastically. So, again, if one places a screen not in the
forward direction parallel to the barrier containing the double slit,
but orthogonally to the latter, one registers an increased local density
of particle arrivals in a narrow spatial area under an angle that
is independent of the slit distance.

\begin{figure}[!tbh]
\subfloat[$a=1$\label{fig:sw.3a}]{\centering{}\includegraphics[width=0.48\textwidth]{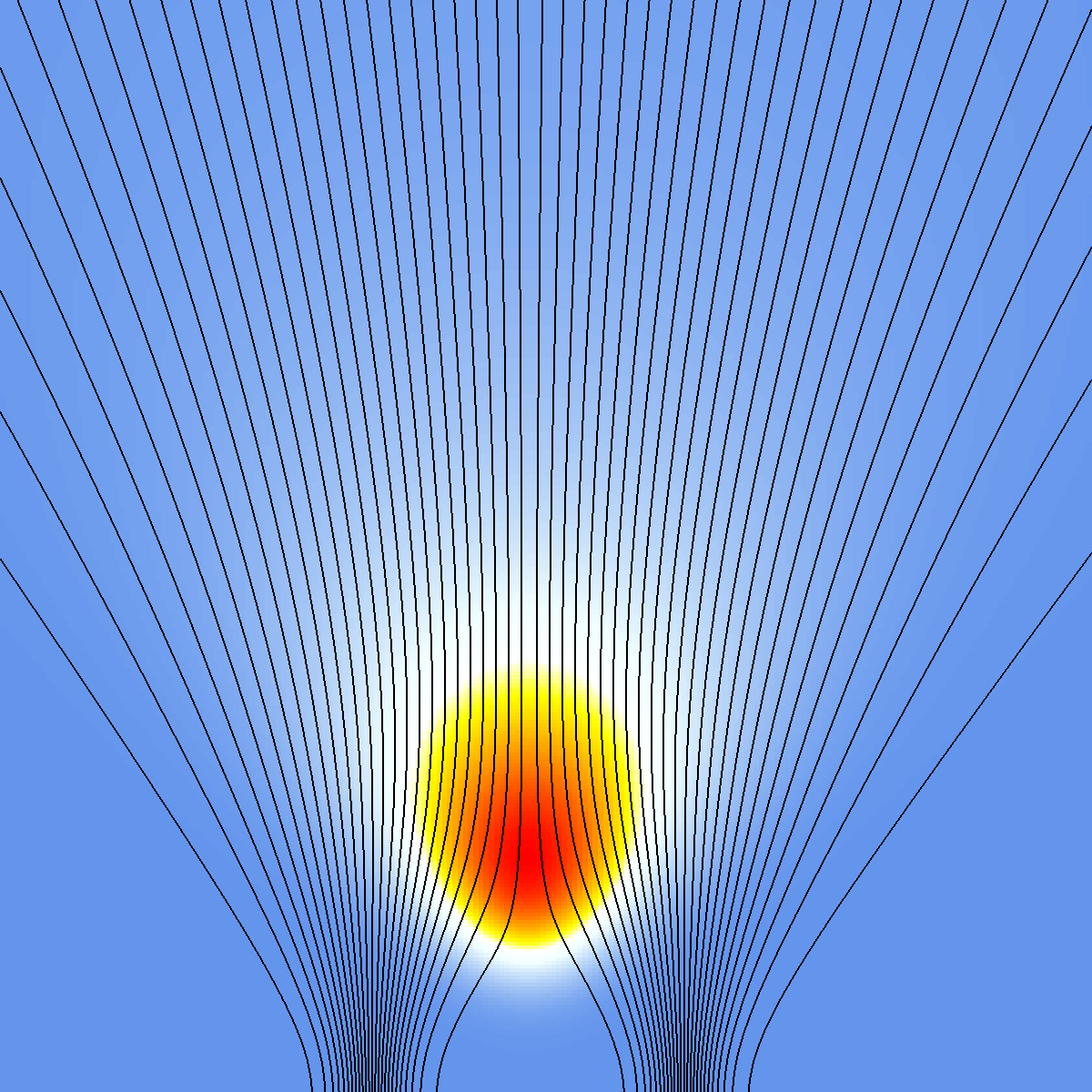}}\hfill{}\subfloat[$a=10^{-8}$\label{fig:sw.3b}]{\centering{}\includegraphics[width=0.48\textwidth]{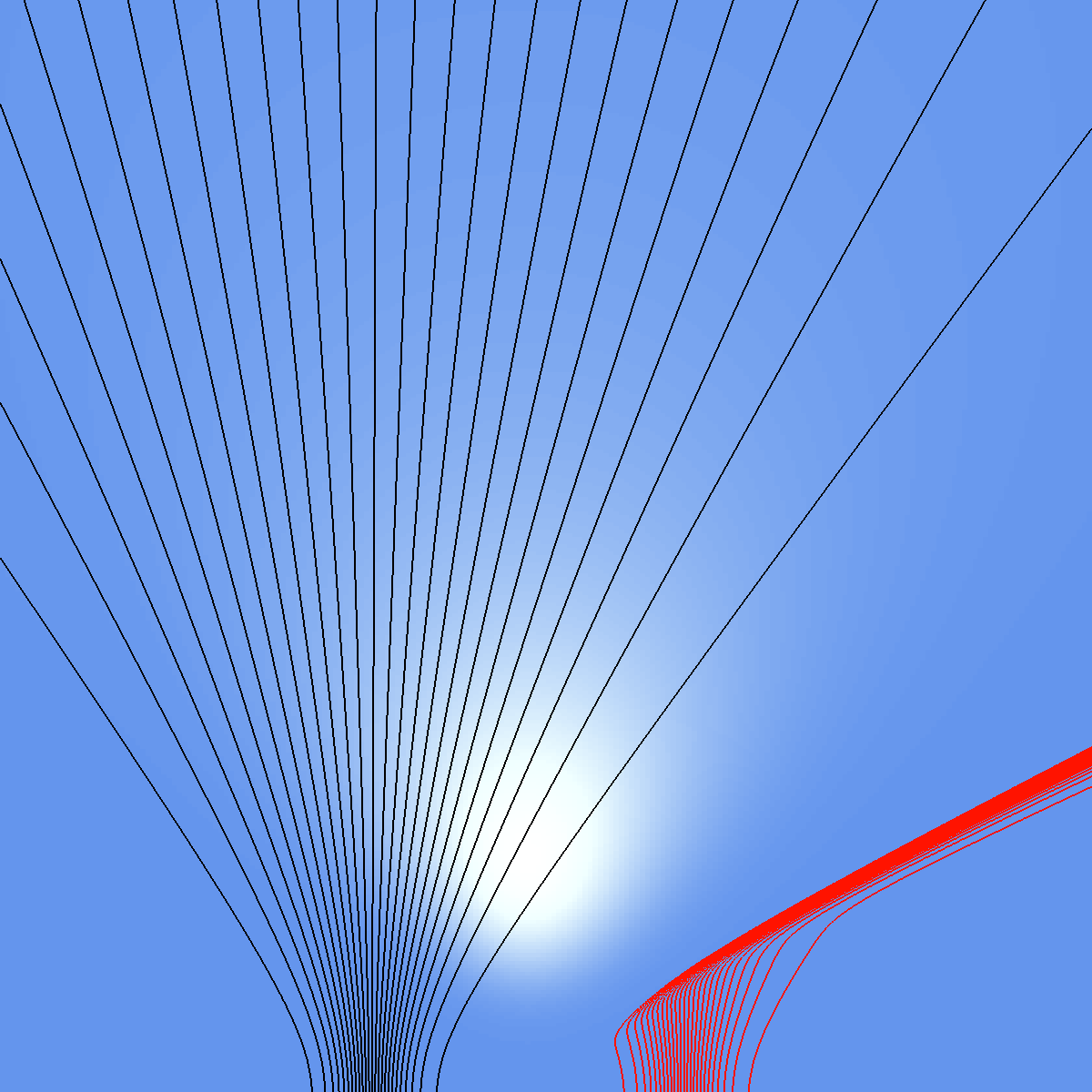}}

\protect\caption{Double-slit experiment with completely incoherent channels. (a) The
average trajectories never cross the central symmetry line, a fact
due to the diffusion related ``hot spot'' indicated in red-to-yellow-to-white
(depicting both interference terms of $J_{x}$), which represents
a kinetic energy reservoir that effectively gives particles a push
in the forward direction. (b) The intensity of $J_{x}$ is now weakened
by the factor $a=10^{-8}$, which is why it does not affect the ``strong''
beam. However, it is sufficient for the attenuated beam to become
deflected.\label{fig:sw.3}}
\end{figure}

Let us now turn to the case of decoherent beams. For, although we
shall refrain from constructing a concrete model of decoherence and
implementing it in our scheme, we already have the tools of an effective
theory, i.e.\ to describe decoherence without the need of a specified
mechanism for it. Namely, as full decoherence between two (Gaussian
or other) beams is characterized by the complete absence of the interference
term in the overall probability distribution of the system, this means
that $P_{\mathrm{tot}}=R_{1}^{2}+R_{2}^{2}$, since the interference
term
\begin{equation}
R_{1}R_{2}\left(\VEC v_{\mathrm{1}}+\VEC v_{2}\right)\cos\varphi=0.\label{eq:sw.15}
\end{equation}
If we therefore choose that on average one has $\cos\varphi=0$, a
situation with $\varphi=\frac{\pi}{2}$ effectively describes two
incoherent beams in the double-slit system. What about the two interference
terms in the probability density current~(\ref{eq:sw.10}), then?
Well, the first term is identical with the vanishing (\ref{eq:sw.15}),
but the second term, with $\VEC u_{i}=-\frac{\hbar}{m}$$\frac{\nabla R_{i}}{R_{i}}$
and $\varphi=\frac{\pi}{2}$ explicitly reads as
\begin{equation}
\frac{\hbar}{m}R_{1}R_{2}\left(\frac{\nabla R_{2}}{R_{2}}-\frac{\nabla R_{1}}{R_{1}}\right)=\frac{\hbar}{m}\left(R_{1}\nabla R_{2}-R_{2}\nabla R_{1}\right).\label{eq:sw.16}
\end{equation}
As the distributions $R_{i}$ may have long wiggly tails -- summing
up, after many identical runs, to a Gaussian with no cutoff, but spreading
throughout the whole domain of the experimental setup~\citep{Groessing.2013dice}
--, the expression~(\ref{eq:sw.16}) is not at all guaranteed to
vanish. In fact, a look at Fig.~\ref{fig:sw.3} shows that there
is an effect even for incoherent beams: Although the product $R_{1}R_{2}$
is negligible and therefore leads to no interference fringes on the
screen, nevertheless expression~(\ref{eq:sw.16}) has the effect
of ``bending'' average trajectories so as to obey the ``no crossing''
rule well known from our model as well as from Bohmian theory.

As was already pointed out in~\citep{Sanz.2009context}, or more
recently, in~\citep{Luis.2013detection}, the resulting trajectories
of Fig.~\ref{fig:sw.3a} can be understood as a nonlinear effect
that is not usually considered in standard quantum mechanics, but
explainable in the Bohmian picture. There, it is the structure of
the velocity field which is genuinely nonlinear and therefore allows
for the emergence of the type of trajectory behavior which has in
earlier years sometimes even been labeled as ``surreal''. However,
also in our approach, the emergence of the trajectories of Fig.~\ref{fig:sw.3}
is completely understandable as it can be traced back to the non-vanishing
of expression~(\ref{eq:sw.16}). Whereas in a naive extrapolation
of what one is used to in standard quantum mechanics, the currents
per channel would -- just like the probability distributions $R_{i}^{2}$
themselves -- go down linearly with ever lowered intensity, this is
actually not the case: It is the nonlinear behavior of the amplitudes
and their gradients, respectively, that are the cause of the observed
trajectory behavior in our simulations.

In sum, then, performing a double-slit experiment with decoherent
beams leads to an emergent behavior of particle propagation which
can be explained by the effectiveness of diffusion waves with velocities
$u_{i}$ interacting with each other, thereby creating a ``hot spot''
where the intensity of the diffusive currents is highest and leads
to a deflection into the forward direction such that no crossing of
the average velocities beyond the symmetry line is made possible (Fig.~\ref{fig:sw.3a}).
This is therefore in clear contradiction to the scenario where only
one slit is open for the particle to go through. If the slits are
not open simultaneously, the particles could propagate to locations
beyond the symmetry line, i.e.\ to locations forbidden in the case
of the second slit being open.~\citep{Sanz.2009context}

As our velocity fields $\VEC v_{i}$ and $\VEC u_{i}$, Eq.~(\ref{eq:sw.12}),
are identical with the Bohmian and the ``osmotic'' momentum, respectively,
one can relate them also to the technique of weak measurements. The
latter have turned out~\citep{Leavens.2005weak,Wiseman.2007grounding,Hiley.2012weak}
to provide said velocities as ``weak values'', which are just given
by the real and complex parts of the quantum mechanical expression
$\frac{\left\langle \mathbf{r}\mid\hat{p}\mid\varPsi\left(\mathbf{\mathit{t}}\right)\right\rangle }{\left\langle \mathbf{r}\mid\varPsi\left(\mathbf{\mathit{t}}\right)\right\rangle }$,
i.e.\ the weak values associated with a weak measurement of the momentum
operator $\hat{p}$ followed by the usual (``strong'') measurement
of the position operator $\hat{r}$ whose outcome is $\mathbf{r}$.
In other words, in principle the trajectories for intensity hybrids
generally, and for the quantum sweeper in particular, are therefore
accessible to experimental confirmation.

Finally, let us stress the relevance of our finding w.r.t.\ the issue
of wave-particle duality. Considering the appearance of compressed
interference fringes in the attenuated beam in Fig.~\ref{fig:sw.2},
it is indisputable that one has to do with the result of a wave-like
behavior. This is confirmed in Fig.~\ref{fig:sw.3b} where the decoherent
scenario is characterized by the complete absence of such wave-like
behavior like interference fringes. This means, however, that an often
used argument to describe the complementarity between wave- and particle-like
behavior in the double slit experiment, or in interferometry, respectively,
has only limited applicability, as it does not apply to intensity
hybrids, since in our model the wave-like contributions due to diffusion
are always present. Specifically, the relation for pure states~\citep{Greenberger.1988simultaneous}
\begin{equation}
D^{2}+V^{2}=1,
\end{equation}
with distinguishability $D=\left|\frac{R_{1}^{2}-R_{2}^{2}}{R_{1}^{2}+R_{2}^{2}}\right|$
representing the particle-related which-path information and visibility
$V=\frac{\mid R_{1}+R_{2}\mid^{2}-\mid R_{1}-R_{2}\mid^{2}}{\mid R_{1}+R_{2}\mid^{2}+\mid R_{1}-R_{2}\mid^{2}}$=$\frac{2R_{1}R_{2}}{R_{1}^{2}+R_{2}^{2}}$
the contrast of the interference fringes in the standard quantum mechanical
double slit scenario suggests that with ever lower values of $a$
ever lower values of $V$ are implied in a constantly decreasing manner.
In our case, by considering the superclassical nature of the sweeper
effect, we find a deviating, characteristic signature at very low
values of $a\lesssim10^{-4}$. In this domain, the usual expectation
would be that practically one has arrived at the ``particle'' side
of the complementarity principle, i.e.\ essentially a one-slit distribution,
with wave-like phenomena having almost disappeared. However, if one
has a very strongly attenuated beam, the emerging behavior of its
outgoing trajectories is different from a one-slit particle distribution
scenario if the other slit is open and un-attenuated. In this case,
the wave-like phenomena do appear, but are restricted solely to the
attenuated beam (Fig.~\ref{fig:sw.4}), a fact that we attribute
to the appearance of nonlinear effects of the probability density
current~(\ref{eq:sw.10}). As already mentioned, the increased local
relative contrast corresponding to a bunching of trajectories and
due to said nonlinear effect of the probability density current is
to be captured by a vertical screen (i.e.\ parallel to the $y$-axis)
for optimal visibility.

In the standard quantum mechanical description of double-slit experiments
with intensity hybrids one is usually only concerned with the gradual
fading out of wave-like properties like interference fringes. However,
in our superclassical model we are dealing with diffusion-based wave-like
properties throughout all magnitudes of attenuation of e.g., slit~2,
even in the case of incoherent beams. For here, if we observe particles
coming through slit~2 characterized by a very low intensity such
as $a=10^{-8}$, one faces the sweeper effect. (Fig.~\ref{fig:sw.4})

The number $n\left(a\right)$ of particles which we do see come through
slit~2 and which produces the distribution (red) in Fig.~\ref{fig:sw.4a}
actually is deflected from the ``forward'' screen when slit~1 is
opened, but the same number $n\left(a\right)$ can easily be detected
on the sideways screen to the right in Fig.~\ref{fig:sw.4b}. Although
the particles would eventually also be detected on a more elongated
forward screen, the effect would be much smaller simply due to the
geometry, whereas the sideways screen setup allows the registration
with maximal contrast. In principle, for beam attenuations as schematized
in Fig.~\ref{fig:sw.4}, if one employs a sideways screen, one thus
obtains a different outcome than the one expected due to standard
quantum mechanical lore. According to the latter, the beam from slit~2
should be unaffected by the situation at slit~1. This would mean
that in the unaffected scenario less than a number of $\frac{n\left(a\right)}{2}$
particles could eventually be registered on any sideways screen parallel
to the $y$-axis along a wide spatial extension, whereas our result
predicts that the totality of the number $n\left(a\right)$ of particles
can be registered within a comparatively narrow spatial domain. In
Fig.~\ref{fig:sw.4c}, the vertical screen setup reveals interesting
features of the probability density distribution, accounting both
for the interference and the sweeper effects. The black line indicates
the continuation of the probability density distribution for the one-slit
case, which is of course being modified once the interference effect
in the coherent case of adding an attenuated beam is allowed for.
However, even in the incoherent scenario not showing the comparatively
small interference effects, one still obtains the full sweeper effect,
with a smooth transition between the two curves in the upper and the
lower parts of Fig.~\ref{fig:sw.4c}, respectively. This is due to
the non-vanishing of (\ref{eq:sw.16}), i.e. a significant contribution
from the diffusive terms despite the smallness of the transmission
factor $a$.

\section{Summary\label{sec:conclusion}}

Summarizing, we have shown that for transmission factors below $a\lesssim10^{-4}$
in intensity hybrids, new effects appear which are not taken into
account in a naive, i.e.\ linear, extrapolation of expectations based
on higher-valued transmission factors. We have described the phenomenology
of these ``quantum sweeper'' effects, including the bunching together
of low counting rate particles within a very narrow spatial domain,
or channel, respectively. However, we also stress that these results
are in accordance with standard quantum mechanics, since we just used
a re-labeling and re-drawing of the constituent parts of the usual
quantum mechanical probability density currents. The reason why the
above-mentioned naive expectations are not met is given by the explicit
appearance of the nonlinear structure of the probability density current
in these domains for very low values of $a$. In this regard, our
subquantum model is better equipped to deal with these appearances
explicitly.

\begin{turnpage}

\end{turnpage}

\begin{ruledtabular}
\end{ruledtabular}
\begin{turnpage}

\end{turnpage}

\appendix

\begin{acknowledgments}
We thank the Fetzer Franklin Fund for partial support of the current
work.
\end{acknowledgments}

\providecommand{\href}[2]{#2}\begingroup\raggedright\endgroup

\end{document}